\documentclass{PNAStwo}

\usepackage{graphicx}
\usepackage{amssymb,amsfonts,amsmath}
\begin{document}



\title{Extracting the hierarchical organization of
complex systems}

\author{Marta Sales-Pardo\affil{1}{Department of Chemical and
Biological Engineering and Northwestern Institute on Complex Systems,
Northwestern University, Evanston, IL 60208, USA}, Roger
Guimer\`a\affil{1}{}, Andr\'e A. Moreira\affil{1}{}
%
%
, and Lu\'{\i}s A. Nunes Amaral\affil{1}{}\thanks{To whom
correspondence should be addressed. E-mail: amaral@northwestern.edu}}

\contributor{Submitted to Proceedings of the National Academy of Sciences
of the United States of America}

\maketitle

\begin{article}

\begin{abstract} Extracting understanding from the
growing ``sea'' of bio\-logical and socio-economic data is one of the
most pressing scientific challenges facing us. Here, we introduce and
validate an unsupervised method that is able to accurately extract the
hierarchical organization of complex biological, social, and
technological networks. We define an ensemble of hierarchically nested
random graphs, which we use to validate the method. We then apply our
method to real-world networks, including the air-transportation
network, an electronic circuit, an email exchange network, and
metabolic networks. We find that our method enables us to obtain an
accurate multi-scale descriptions of a complex system.
\end{abstract}


\dropcap{T}he high-throughput methods available for probing biological
samples have drastically increased our ability to gather comprehensive
molecular-level information on an ever growing number of organisms.
These data show that these systems are connected through a dense
network of nonlinear interactions among its
components\:\cite{anderson74,hartwell99}, and that this
interconnectedness is responsible for their efficiency and
adaptability. At the same time, however, such interconnectedness poses
significant challenges to researchers trying to interpret empirical
data and to extract the ``systems biology'' principles that will
enable us to build new theories and to make new predictions
\cite{pennisi05}.

A central idea in biology is that life processes are hierarchically
organized \cite{hartwell99,oltvai02,alon03,itzkovitz05} and that this
hierarchical structure plays an important role in their dynamics
\cite{arenas06}. However, given a set of genes, proteins, or
metabolites and their interactions, we still do not have an objective
manner to assess whether such hierarchical organization does indeed
exist, or to objectively identify the different levels in the
hierarchy.

Here, we report a new method that identifies the levels in the
organization of complex systems and extracts the relevant information
at each level.  To illustrate the potetial of our method, it is useful
to think of electronic maps as in {\sf \small http://maps.google.com}
(Fig.\:S1). Electronic maps are powerful tools because they present
information in a scalable manner, that is, despite the increase in the
amount of information as we ``zoom out,'' the representation is able
to extract the information that is relevant at the new scale.  In a
similar spirit, our method will enable researchers to characterize
each scale with the relevant information at that scale. This
achievement is key for the development of systems biology, but will
encounter application in many other areas.

\section*{Background}

Complex networks are convenient representations of the interactions
within complex systems\:\cite{amaral04}.
Here, we focus on the identification of inclusion hierarchies in
complex neworks, that is, to the unraveling of the nested organization
of the nodes in a network into modules, which are comprised of
sub-modules and so on\footnote{We do not consider other hierarchical
schemes that classify nodes according to, for instance, their
importance\:\cite{trusina04}. Another issue that we do not address here
is that of ``overlapping'' modules.  In the literature, some authors
refer to the existence of ``soft'' boundaries between
communities\:\cite{reichardt04,ziv05}. However, there has been so far
no rigorous connection between the soft boundaries and the overlap
between communities. Moreover, at present, there is no theoretical
model that includes overlapping modules, that is, modules that share
nodes, as opposed to communities that share edges.}.

A method for the identification of the hierarchical organization of
nodes in a network must fulfill two requirements: (i) it must be
accurate for many types of networks, and (ii) it must identify the
different levels in the hierarchy as well as the number of modules and
their composition at each level. The first condition may appear as
trivial, but we make it explicit to exclude algorithms that only work
for a particular network or family of networks, but that will
otherwise fail. The second condition is more restrictive, as it
excludes methods whose output is subject to
interpretation. Specifically, a method does not fulfill the second
condition if it organizes nodes into a tree structure, but it is up to
the researcher to find a ``sensible'' criterion to establish which are
the different levels in that tree. An implication of the previous two
requirements is that any method for the identification of node
organization must have a null output for networks, such as
Erd\H{o}s-R\'enyi random graphs, which do not have an internal
structure.

To our knowledge, there is no procedure that enables one to
simultaneously assess whether a network is organized in a hierarchical
fashion and to identify the different levels in the hierarchy in an
unsupervised way. Ravasz et al. \cite{ravasz02} studied the
hierarchical structure of metabolic networks, but in their analysis
the authors put emphasis on detecting ``global signatures'' of a
hierarchical network architecture. Specifically, they reported that,
for the metabolic networks studied and for certain hierarchical
network models, the clustering coefficient of nodes appears to scale
with the connectivity as $C(k)\sim k^{-1}$. This scaling, however, is
neither a necessary nor a sufficient condition for a network to be
hierarchical \cite{soffer05}.

More direct methods to investigate the hierarchical organization of
the nodes in a network have also been recently proposed
\cite{guimera03,clauset06,pons06}. Although useful in some contexts,
these methods do not clearly identify hierarchical levels and thus
fail to satisfy condition (ii) above. Furthermore, all these methods
yield a tree even for networks with no internal structure.

In the following, we define inclusion hierarchies in complex networks
and describe an ensemble of hierarchically nested random graphs. We
then introduce a method that is able to accurately extract the
hierarchical organization of hierarchical random graphs. Finally, we
apply our method to several real-world networks.

\section*{Inclusion hierarchies}


Consider the ensemble of networks comprised of $N$ nodes, $\mathcal{N}
=$ \{$\,n_i: i = 1, \dots, N\,$\}, that hold membership in a set of
nested groups, $\mathcal{G}=$\{$\,g_{(k_1\:\dots\:k_\ell)}:
\ell=1,2\dots\,$\}, where $\ell$ is the level at which the group is
defined, and the labels $k_1\:\dots\:k_{\ell-1}$ indicate the groups
at higher levels in the hierarchy within which the group is
nested. For instance, group $g_{111}$ is a group defined at $\ell=3$
that is nested inside group $g_{11}$ defined at $\ell=2$, which in
turn is a subgroup of group $g_1$ defined at $\ell=1$.


Let $\mathcal{G}_i\subset \mathcal{G}$ be the set of groups in which
node $n_i$ holds membership. Here, we consider that node $n_i$ holds
membership in only one group per level, and that membership to groups
follows a nested hierarchy. Therefore, for node $n_i$ to hold
membership in group $g_{11}$, node $n_i$ must also hold membership in
group $g_{1}$.

We assume that the probability $p_{ij}$ of the edge $(n_i, n_j)$ being
present in a network is a function {\it solely\/} of the set of
co-memberships $\mathcal{M}_{ij} = \mathcal{G}_{i} \cap
\mathcal{G}_{j}$ of the two nodes. Note that our assumptions imply
that: (i) $\mathcal{M}_{ij}$ obeys transitivity, so that if
$\mathcal{M}_{ij} = \mathcal{M}_{ik} $, then $\mathcal{M}_{ij} =
\mathcal{M}_{jk}$; and (ii) node memberships in groups
\{$\,g_{k_1k_2}\,$\} at the second level are uniquely and completely
defined by the sub-network of connections of all nodes holding
membership in group $g_{k_1}$, that is, information at deeper levels
in the hierarchy is totally decoupled from the information at higher
levels in the hierarchy.

In the simplest scenario, $p_{ij}$ is a non-decreasing function of the
cardinality $x$ of $\mathcal{M}_{ij}$, which implies that groups of
nodes holding membership in the same groups will be more densely
connected than a randomly selected group of nodes. This is precisely
the underlying assumption in many algorithms aiming to detect the top
level community structure of complex networks assuming a flat
organization of the nodes\:\cite{newman04,guimera05a,duch05}.

Let us now introduce an ensemble of random networks which are
constructed following hierarchical node membership assignment:
hierarchically nested random graphs. We restrict our ensemble to
networks with a homogeneous hierarchical organization of the nodes
(see Supplementary Information for other kinds of hierarchical
organization) that have the same degree distribution as
Erd\H{o}s-R\'enyi graphs \cite{watts99}.

To illustrate the model, consider a network comprised of 640 nodes
that hold membership in a set of groups $\mathcal{G}$ with a
three-level homogeneous nested organization. We assign group
memberships so that the number $S_\ell$ of nodes holding membership in
each group for $\ell=1,2$, and $3$ is $S_1=160$, $S_2=40$, and
$S_3=10$, respectively. For $\ell=1$, nodes can hold membership in one
of four different groups \{$\,g_{k_1}:\:k_1=1,\dots,4\,$\}. For
$\ell=2$, nodes holding membership in group $g_{k_1}$ can hold
membership in one of four groups \{$\,g_{k_1k_2}:
k_2=1,\dots,4\,$\}. Finally, for $\ell=3$, nodes holding membership in
groups $g_{k_1}$ and $g_{k_1k_2}$ can hold membership in one of four
groups \{$\,g_{k_1k_2k_3} : k_3=1,\dots,4\,$\}.

The probability $p_x$ of edge $(n_i, n_j)$ existing is a monotonically
growing function that depends exclusively on the cardinality $x$ of
$\mathcal{M}_{ij}$. Thus, if the expected number of links between
$n_i$ and nodes \{$\{n_j\} : ||\mathcal{M}_{ij}|| = x$\} is $k_x = p_x
S_x$. Probabilities are chosen so that the average degree of a node is
$\overline{k} = \sum_{\ell=0}^{\ell_{\rm max}} \overline{k_\ell}$, and
the ratio $\rho = \overline{k_{<l}}/\overline{k_{l}}$ is constant
throughout the levels, where $\overline{k_{<\ell}} =
\sum_{\ell'=0}^{\ell-1}
\overline{k_{\ell'}}$.\:\footnote{For example, for the three-level
network described earlier, and $\overline{k}=16$ and $\rho=1$,
$\overline{k_0}=8$, $\overline{k_1}=4$, $\overline{k_2}=2$, and
$\overline{k_3}=3$ (see Supplementary Material for the expression of
$p_x$).} The reason for such choice is to facilitate both the graphic
representation and the interpretation of the results. Note that, for
$\rho<1$, deeper levels are more cohesive, whereas for $\rho>1$, they
are less cohesive (Supplementary Information).


%
\section*{Extracting the hierarchical organization of networks}

Our method consists of two major steps (Fig.\:\ref{f.steps}): (i)
measuring the ``proximity'' in the hierarchy between all pairs of
nodes, which we call {\it node affinity}; and (ii) uncovering the
overall hierarchical organization of node affinities, or, in other
words, detecting the underlying organization of group memberships.

\subsection*{Node affinity---}

A standard approach for quantifying the affinity between a pair of
nodes in a network is to measure their ``topological overlap''
\cite{ravasz02,ye04,ihmels05}, which is defined as the ratio between
the number of common neighbors of the two nodes and the minimum degree
of the two nodes. This measure identifies affinity between nodes with
a dense pattern of local connections. Because topological overlap is a
local measure, it will fail to detect any structure when a network is
not locally dense (Fig.\:\ref{f.comparison}).

We propose a new affinity measure based on surveying of the modularity
landscape\:\cite{newman04b}, a collective property of the network. Our
definition of affinity between nodes draws upon the idea that modules
correspond to sets of nodes which are more strongly interconnected
than one would expect from chance alone \cite{newman04b,clauset05}. We
show below that our affinity measure detects the modular structure
even in the absence of a dense pattern of local connections.

Consider the ensemble $\cal P$ of all partitions of a network into
modules \cite{newman04b,guimera04}, and assign to each partition $P$
the modularity
%
%
\begin{equation}
M(P) = \sum_{i=1}^m\left[\frac{l_i}{L}-\left(\frac{d_i}{2L}\right)^2\right]~~,
\end{equation} 
%
where $L$ is the total number of links in the network, $l_i$ is the
number of links within module $i$, $d_i$ is the sum of degrees of all
the nodes inside module $i$, and the sum is over all the $m$ modules
in partition $P$ (Fig.\:\ref{f.steps}A). The modularity of a partition
is high when the number of intra-module links is much larger than
expected for a random partition.

Let $\cal P_{\rm max}$ be the set of partitions for which the
modularity $M$ is a local maxima, that is, partitions for which
neither the change of a single node from one module to another nor the
merging or splitting of modules will yield a higher modularity
\cite{guimera05}. Let $B_{\rm max}=\{\,b(P):P\in\cal P_{\rm max}\,\}$
be the sizes of the ``basin of attraction'' of those maxima. The
affinity $A_{ij}$ of a pair of nodes $(i,j)$ is then the probability
that when local maxima $P\in \cal P_{\rm max}$ are sampled with
probabilities $b(P)$, nodes $(i,j)$ are classified in
the same module.

Note that, in contrast to other affinity measures proposed in
Refs.\:\cite{ziv05,pons06,newman04b}, the measure we propose does not
necessarily coincide with the ``optimal'' division of nodes into
modules, that is, the partition that maximizes
$M$\:\cite{fortunato07}. In fact, the modules at the top level of the
hierarchy do not necessarily correspond to the best partition found
for the global network, even for relatively simple networks
(Fig.\:\ref{f.comparison}C).


\subsection*{Statistical significance of hierarchical organization---}
Given a set of elements and a matrix of affinities between them, a
commonly used tool to cluster the elements and, presumably, uncover
their hierarchical organization is hierarchical clustering
\cite{everitt01,schena02}. Hierarchical clustering methods have three
major drawbacks: (i) They are only accurate at a local level---at
every step a pair of units merge and some details of the affinity
matrix are averaged with an inevitable loss of information; (ii) the
output is always a hierarchical tree (or dendogram), regardless of
whether the system is indeed hierarchically organized or not; (iii)
there is no statistically sound general criterion to determine the
relevant levels on the hierarchy.

In order to overcome the first caveat of agglomerative methods such as
hierarchical clustering, one necessarily has to follow a top to bottom
approach that keeps the details of the matrix. That is the spirit of
divisive methods such as k-means or principal component
analysis\:\cite{everitt01}, which group nodes into ``clusters'' given
an affinity matrix. However, these methods have a significant
limitation: the number of clusters is an external parameter, and,
again, there is no sound and general criterion to objectively
determine the correct number of clusters.

Because of the caveats of current agglomerative and divisive methods,
we propose a ``box-clustering'' method that iteratively identifies in
an unsupervised manner the modules at each level in the hierarchy.
Starting from the top level, each iteration corresponds to a different
hierarchical level (Fig.\:\ref{f.comparison}).

In order to assess whether the network under analysis has an internal
organization we need to compare with the appropriate null model, which
in this case is an ensemble of ``equivalent'' networks with no
internal organization. These equivalent networks must have the same
number of nodes and an identical degree sequence. A standard method
for generating such networks is to use the Markov-chain switching
algorithm\:\cite{maslov02,guimera07}. Despite their having no internal
structure, these networks have numerous partitions with non-zero
modularity\:\cite{guimera04}. Thus, to quantify the level of
organization of a network, one needs to compare the modularities of
the sampled maxima for the original network and its corresponding
random ensemble; if the network has a non-random internal structure,
then local maxima in the original landscape should have larger
modularities than local maxima in the landscapes of the randomized
networks.

Specifically, for a given network, we compute the average modularity
$M_{\rm av}$ from \{\:$M(P)\::\:P \in \cal P_{\rm max}\:$\}. Then, we
compute the same quantity $M^i_{\rm av}$ for each network in the
equivalent random ensemble. In virtue of the central limit theorem,
the set of average modularities for the whole ensemble \{$\:M^i_{\rm
av}\:$\} is normally distributed with mean $M_{\rm rand}$ and variance
$\sigma^2_{M_{\rm rand}}$. To quantify the level of organization of a
network, we thus compute the z-score of the average modularity
\begin{equation}
z = \frac{ M_{\rm av} - M_{\rm rand} }{ \sigma_{M_{\rm rand}} }~.
\end{equation}
If $z$ is larger than a threshold value $z_t$, then the network has
internal structure and we proceed to identify the different modules,
otherwise we conclude that the network has no structure. In what
follows, we show results for $z_t = 2.3267$, which corresponds to a
1\% significance level (Supplementary Material)\footnote{Results for
real networks at a 5\% significance level are identical, however, the
more stringent threshold is more efficient at detecting the last level
in the hierarchy for model networks. Only for a 1-3\% of the
cases---depending on the cohesiveness of the levels---do we find that
algorithm finds one more level than expected.}.

\subsection*{Building the hierarchical tree---}
 
\begin{figure}[t]
\center \includegraphics*[width=1\columnwidth]{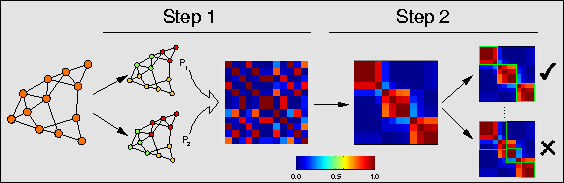}
\caption{Schematic illustration of our method.
Step~1: Affinity matrix. Sampling of the maxima of the of the
modularity landscape. We use the co-classification of nodes in the
same module for partitions that are a local maxima of the modularity
landscape as a measure of the affinity between the nodes. We then
verify whether the network has a non-random internal structure. If it
does not, we stop here.
Step~2a: Ordering the affinity matrix and extraction of the
hierarchical organization. The affinity matrix will show a
hierarchical organization of the nodes, if pairs of nodes with high
affinities occupy contiguous rows in the matrix. To find the optimal
ordering of the nodes, we define a ``cost function'' that weighs
each matrix element by its distance to the diagonal.
Step~2b: Extracting the hierarchical organization. The signature of a
hierarchical organization is the existence of a nested block diagonal
structure in the affinity matrix. In order to identify the different
modules (boxes) at each level $\ell$ in the hierarchy, we propose an
ansatz matrix with $n$ boxes of identical elements along the diagonal
$A^s_\ell$, for $s=0,...,n$, and identical elements $B_\ell$ outside
the boxes.  We use a ``least-squares'' method combined with a ``greedy
algorithm'' to determine the partition that best fits the model (see
text and Supplementary Information). We go back to step 1a for each
one of the sub-networks defined by the partition. 
}
\label{f.steps}
\end{figure}

\begin{figure}
\center \includegraphics*[width=\textwidth]{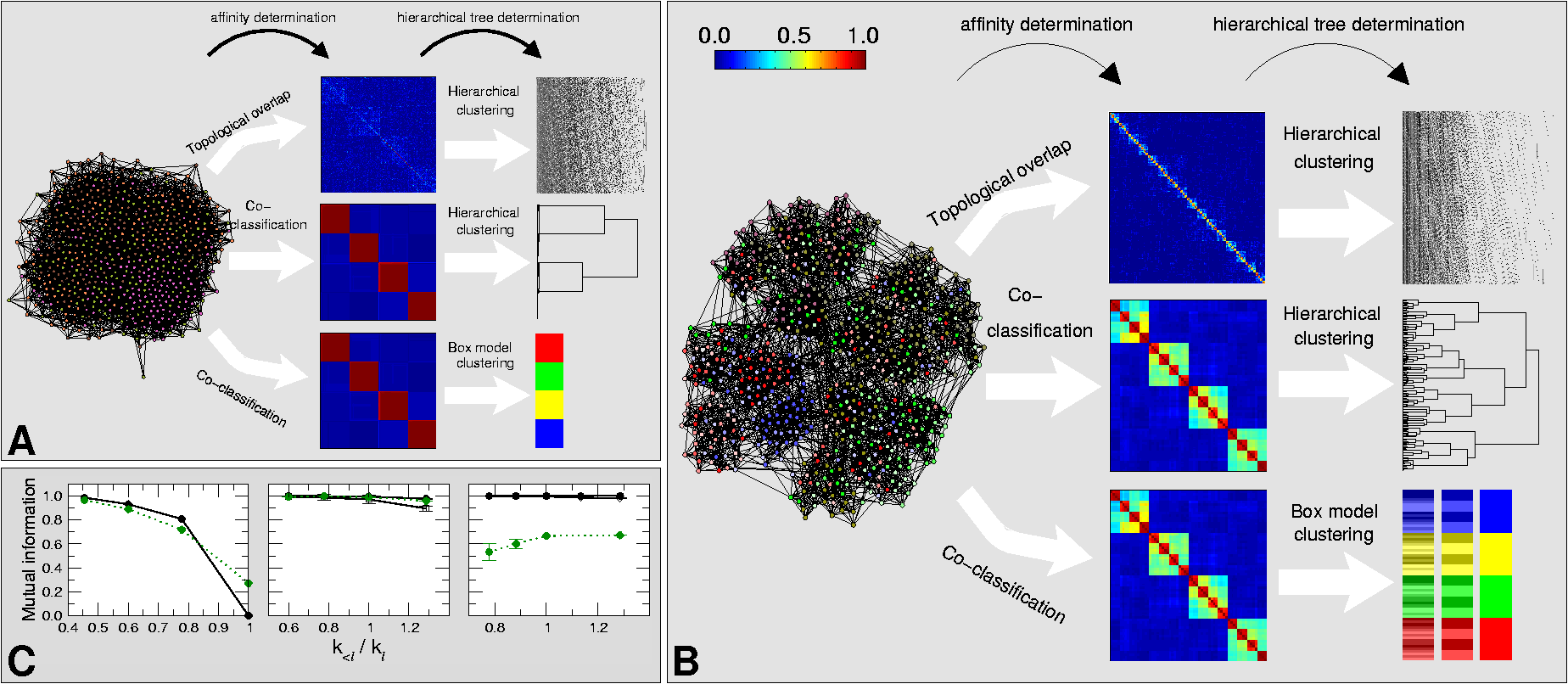}

\caption{Affinity measures and clustering methods.
We generate two model networks comprised of 640 nodes with average
degree 16.  {\bf A,} Modular network with ``flat'' structure. The
network comprises four modules with 160 nodes each.  The nodes have an
average of eleven within-module connections and five inter-module
connections; {\bf B,} Modular network with a three-level hierarchical
structure.
We show affinity matrices $A_{ij}$ obtained for two different
measures: (i) topological overlap\:\cite{ravasz02}; (ii)
``co-classification''(see text and Supplementary Information). The
color scale goes from red for a probability of one to dark blue for a
probability of zero.
At the far right, we show the hierarchical tree obtained using two
different methods: hierarchical clustering and the ``box
clustering'' we propose. In the hierarchical clustering tree, the
vertical axis shows the average distance,
$\overline{d_{ij}}=\overline{1-A_{ij}}$, of the matrix elements that
have already merged. In the box-model clustering tree, each row
corresponds to one hierarchical level. Different colors indicate
different modules at that level. To better identify which are the
sub-modules at a lower level, we color the nodes in the sub-modules
with shades of the color used for the modules in the level above.
Note that topological overlap fails to find any modular structure
beyond a locally dense connectivity pattern. In contrast, the
co-classification measure clearly reveals the hierarchical
organization of the network by the ``nested-box'' pattern along the 
diagonal.
Significantly, the hierarchical tree obtained via hierarchical
clustering fails to reproduce the clear three-level hierarchical
structure that the affinity matrix displays, whereas the
box-model clustering tree accurately reproduces the three-level
hierarchical organization of the network.
{\bf C,} Accuracy of the method.
We generate networks with 640 nodes and with built-in hierarchical
structure comprising one (left), two (middle), and three (right)
levels. The top level always comprises four modules of 160 nodes each.
For networks with a second level, each of the top-level modules is
organized into four sub-modules of 40 nodes.  For the networks with
three levels, each level-two module is further split into four
sub-modules of ten nodes. We build networks with different degrees of
level cohesiveness by tuning a single parameter $\rho$ (see text). 
%
Since we know {\em a priori} which are the nodes that should be
co-classified at each level, we measure the accuracy as the mutual
information between the empirical partition of the nodes and the
theoretical one\:\cite{danon05}. We also plot the accuracy of a
standard community detection algorithm\:\cite{newman06} in finding the
top level of the networks (dashed green line).
We plot the mutual information versus $\rho$ for networks with one
(left), two (center) and three (right) hierarchical levels. Each point
is the average over ten different realizations of the network. Full
circles, empty squares, and full diamonds represent the accuracy at
the top, middle, and lowest levels, respectively. 
}
\label{f.comparison}
\end{figure}

In networks organized in a hierarchical fashion, nodes that belong to
the same module at the bottom level of the hierarchy have greater
affinity than nodes that are together at a higher level in the
hierarchy. Thus, if a network has a hierarchical organization, one
will be able to order the nodes in such a way that groups of nodes
with large affinity are close to each oder. With such an ordering, the
affinity matrix will then have a ``nested'' block-diagonal structure
(Fig.\:\ref{f.steps}). This is indeed what we find for networks
belonging to the ensemble of hierarchically nested random graphs
(Fig.\:\ref{f.comparison}).

For real-world networks, we do not know {\em a priori} which nodes are
going to be co-classified together, that is, we do not know which is
the ordering of the nodes for which the affinity matrix has a nested
block-diagonal structure. To find such an ordering, we use simulated
annealing \cite{kirkpatrick83} to minimize a cost function that weighs
each matrix element with its distance to the diagonal
\cite{wasserman94}
%
\begin{equation}
{\cal C}=\frac{1}{N}\sum_{i,j=1}^{N} A_{ij}|i-j|, 
\end{equation}
%
where $N$ is the order of the affinity matrix (see Fig.\:\ref{f.steps}A
and Supplementary Information for alternative ordering schemes).

This problem belongs to the general class of quadratic assignment
problems\:\cite{koopmans75}. Other particular cases of quadratic
assignment problems have been suggested to uncover different features
of similarity matrices\:\cite{tsafrir05}
Our algorithm is able to find the proper ordering for the affinity
matrix and to accurately reveal the structure of hierarchically nested
random graphs (Fig.\:\ref{f.comparison}).

\paragraph{Unsupervised extraction of the structure---} 

Given an ordered affinity matrix, the last step is to partition the
nodes into modules at each relevant hierarchical level. An {\it
ansatz\/} that follows naturally from the considerations in the
previous section and the results in Fig.\:\ref{f.comparison} is that,
if a module at level $\ell$ (or the whole network at level 0) has
internal modular structure, the corresponding affinity matrix is
block-diagonal: At level $\ell$, the matrix displays boxes along the
diagonal, such that elements inside each box $s$ have an affinity
$A^s_\ell$, while matrix elements outside the boxes have an affinity
$B_\ell < A^s_\ell$. Note that the number of boxes for each affinity
matrix is not fixed; we determine the ``best'' set of boxes by least
squares fitting of the block-diagonal model to the affinity matrix.

Importantly, we want to balance the ability of the model to accurately
describe the data with its parsimony, that is, we do not want to
over-fit the data. Thus, we use the Bayesian information criterion in
order to determine the best set of boxes\:\cite{schwarz78}\:\footnote{We
have also applied Akaike's information criterion\:\cite{akaike74},
obtaining the same results for most of the cases.}.



To find the modular organization of the nodes at the top level (level
1), we fit the block diagonal model to the global affinity matrix. As
we said previously, we assume that the information at different levels
in the hierarchy is decoupled, thus in order to detect sub-modules
beyond the first level, one needs to break the network into the
sub-networks defined by each module and apply the same procedure
(Fig.\:\ref{f.steps}).
%
 The algorithm iterates these steps for each identified box
until no sub-networks are found to have internal structure.

\section*{Method validation}

We validate our method on hierarchically nested random graphs with
one, two, and three hierarchical levels. We define the accuracy of the
method as the mutual information between the empirical partition and
the theoretical one \cite{danon05}.  Figure\:\ref{f.comparison}C shows
that the algorithm uncovers the correct number of levels in the
hierarchy.


Moreover, our method always detects the top level, even for the
networks with three hierarchical levels. In contrast, because the
partition that globally maximizes $M$ corresponds to the sub-modules
in the second level, even the more accurate module identification
algorithms based on modularity maximization would fail to capture the
top level organization (Joshi {\em et al.} 2007,\:\cite{fortunato07}).

The hierarchically nested random graphs considered above have a
homogeneous hierarchical structure; however, real-world networks are
not likely to be so regular. In particular, for real-world networks
one expects that some modules will have deeper hierarchical structures
than others. We thus have verified that our method is also able to
correctly uncover the organization of model networks with
heterogeneous hierarchical structures (Supplementary Information).

\section*{Testing on real world networks} 

Having validated our method, we next analyze different types of
real-world networks for which we have some insight into the network
structure: the world-wide air-transportation network
\cite{barrat04,guimera04a,guimera05b}, an e-mail exchange network of a
Catalan university \cite{guimera03}, and an electronic circuit
\cite{itzkovitz05}.

In the air transportation network, nodes correspond to airports and
two nodes are connected if there is a non-stop flight connecting
them. In the email network, nodes are people and two people are
connected if they send emails to each other. In the electronic
network, nodes are transistors and two transistors are connected if
the output of one transistor is the input of the other
(Table\:\ref{t.realnetworks}).

\begin{figure}
\center \includegraphics*[width=0.8\textwidth]{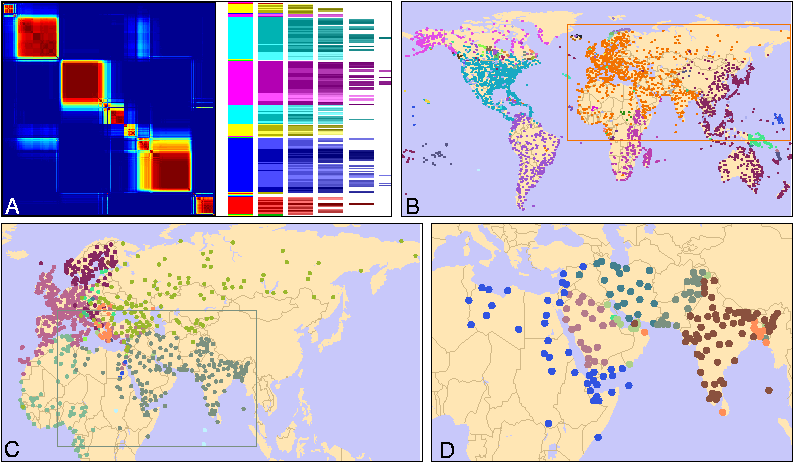}
\caption{Hierarchical organization of the air-transportation network.
{\bf A,} Global-level affinity matrix and hierarchical tree (the
representation is the same used in Fig.\:\protect\ref{f.comparison}).
{\bf B,} Top-level modules. Each dot represents an airport and
different colors represent different modules. Note that the top level
in the hierarchy corresponds roughly to geo-political units.
The ``orange'' module (comprised of the majority of European countries,
ex-Soviet Union countries, Middle-Eastern countries, India, and
countries in Northern half of Africa) splits for levels $\ell=2$ ({\bf
C}) and $\ell=3$ ({\bf D}).
%
}
\label{f.airports}
\end{figure}

We find that the air-transportation network is strongly modular and
has a deep hierarchical organization (Fig.\:\ref{f.airports}). This
finding does not come as a surprise since historical, economic,
political, and geographical constraints shape the topology of the
network \cite{barrat04,guimera04a,guimera05b}. We find eight main
modules that closely match major continents and sub-continenets, and
major political divisions and thus truly represent the highest level
of the hierarchy\footnote{The ability of the present method to detect
the top level is significant. A previous study co-authored by two of
us identified 19 modules in the world-wide air-transportation network
\cite{guimera05b} using the most accurate module detection algorithm
in the literature \cite{guimera05a}.}.

\begin{figure}
\center \includegraphics*[width=0.9\columnwidth]{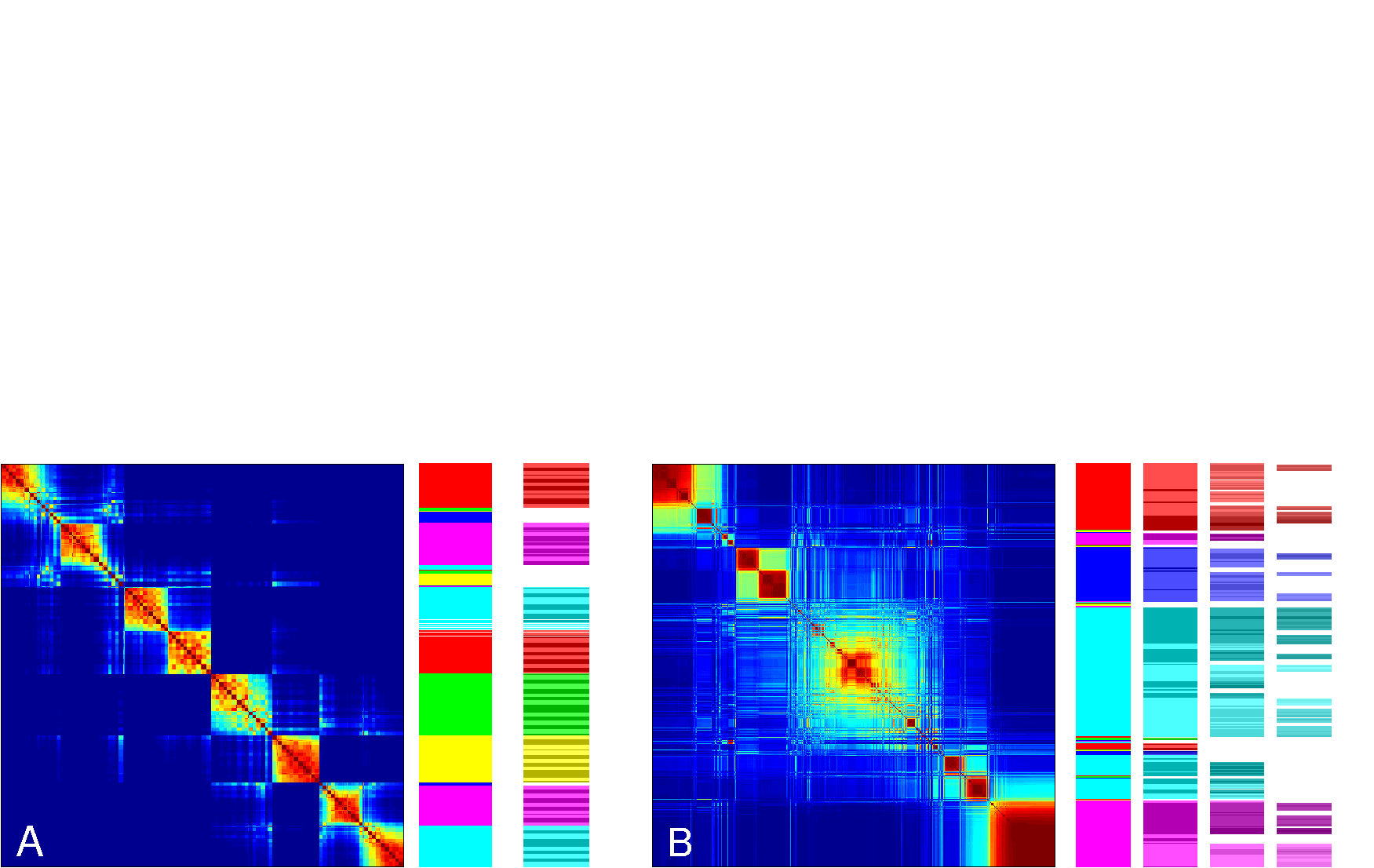}
\caption{Hierarchical structure of technological and social
networks.
We show the ordered affinity matrices at the top level and the
hierarchical trees that we obtain for {\bf A,} the transistor
implementation of an electronic circuit \cite{itzkovitz05}, and {\bf
B,} the email exchange network of a Catalan university
\cite{guimera03}.
Our method is capable of accurately uncovering the top level
organization of the networks. For the transistor network, which is
comprised of eight D-type flipflops and 58 logic gates, we find that
at the top level, gates comprising a given D-flipflop are classified
in the same module. A the second level, the majority of the modules
are comprised of a single gate.  For the email network, at the top
level we find eight modules that closely match the organization of the
schools and centers in the university \cite{guimera03}.
}
\label{f.email-tecno}
\end{figure}

The electronic circuit network is comprised of eight D-flipflops and
58 logic gates \cite{itzkovitz05}. Our method identifies two levels in
the network (Fig.\:\ref{f.email-tecno}A). At the top level, modules
are groups of logic gates, all the logic gates comprising a D-flipflop
being in the same module. At the second level, the majority of modules
comprise single gates.

For the email network, five of the seven major modules at the top
level (Fig.\:\ref{f.email-tecno}B) correspond to schools in the
university, with more than 70\% of the nodes in each of those modules
affiliated to the corresponding school. The remaining two major
modules at the top level are a mixture of schools and administration
offices (often collocated on campus), which are distinctly separated
at the second level. The second level also identifies major
departments and groups within a school, as well as research centers
closely related to a school.

\section*{Application to metabolic  networks}

\begin{figure}[b]
\center \includegraphics*[width=.5\columnwidth]{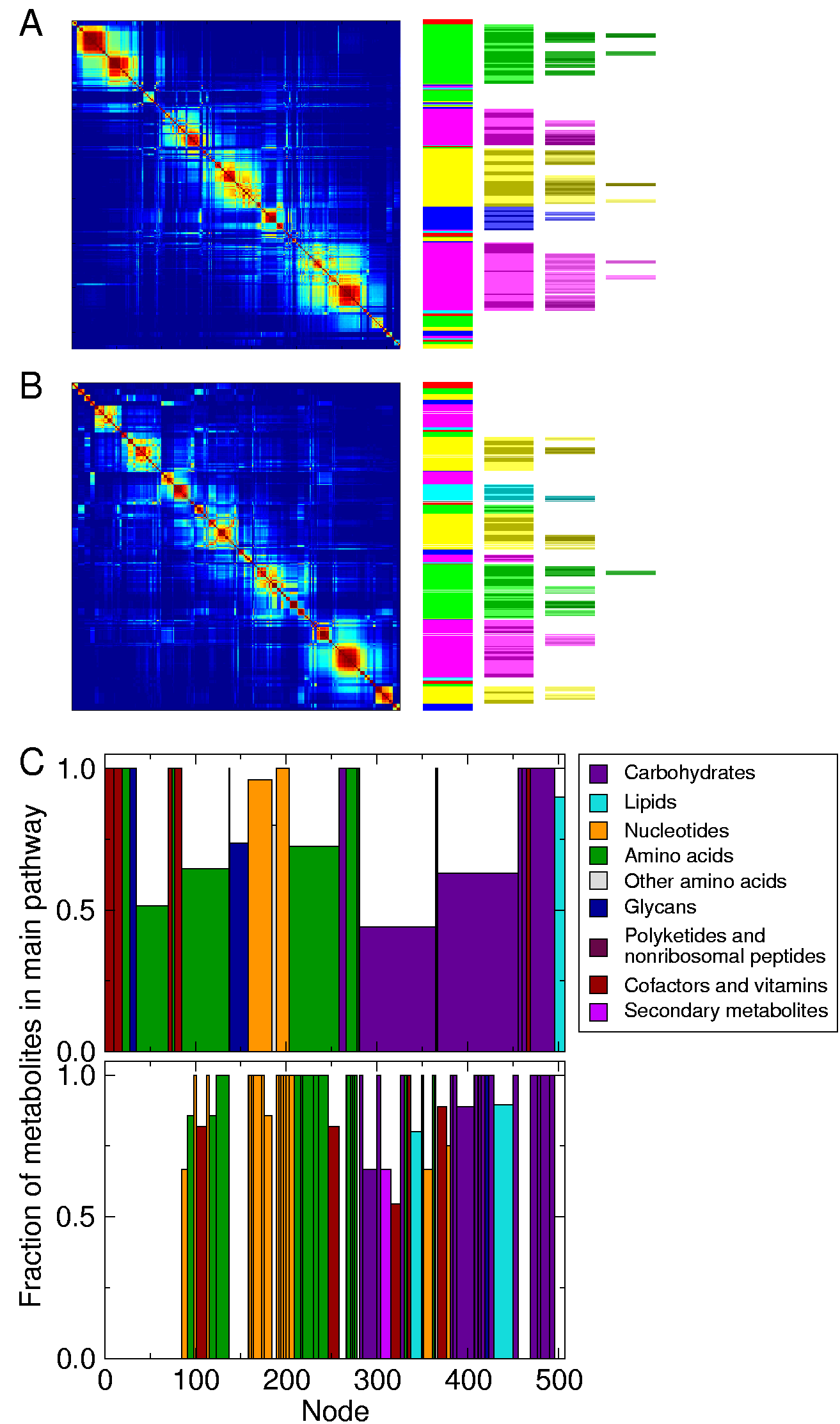}
%
%
\caption{Hierarchical structure of metabolic networks.
Global level affinity matrices and hierarchical trees for the
metabolic networks of {\em E. coli} obtained from: {\bf A,} the KEGG
database \cite{goto98,kanehisa00}, and {\bf B,} the Systems Biology
group at UCSD \cite{reed03}.
Note that the overall organization of the networks is similar and
independent of the reconstruction used to build the network.
{\bf C,} For the metabolic network of {\it E. coli} from the Systems
Biology group at UCSD, we analyze the within-module consistency of
metabolite pathway classification for the first (top plot) and the
second (bottom plot) levels. For each module, we first identify the
pathway classifications of the corresponding metabolites; then, we
compute the fraction of metabolites that are classified in the most
abundant pathway.
In the plots, each bar represents one module, its width being
proportional to the number of nodes it contains. We color each bar
according to its most abundant pathway following the color code on the
right hand side. At the second level (bottom plot), we show each
sub-module directly below its corresponding top level module. Again,
the width of each sub-module is proportional to its size.
Note that, at the first level (top), for all modules except one, the
most abundant pathway is comprised of more than 50\% of the
metabolites in the module. Remarkably, at the second level (bottom),
for most of the modules all the metabolites are classified in the same
pathway.
Moreover, at the second level, we detect smaller pathways that are not
visible at the top level.}
\label{f.metabolic}
\end{figure}

Finally, we analyze the metabolic networks of {\it E. coli\/} obtained
from two different sources\footnote{In the Supplementary Material we
also show the organization obtained for the metabolic network for {\em
E. coli} from the Ma-Zeng database \cite{ma03}, and for the metabolic
network of {\em H. pylori} developed at UCSD \cite{thiele05}.}
(Fig.\:\ref{f.metabolic}): the KEGG database \cite{goto98,kanehisa00},
and the reconstruction compiled by Palsson's Systems Biology Lab at
UCSD \cite{reed03}. In these networks, nodes are metabolites and two
metabolites are connected if there is a reaction that transforms one
into the other\:\cite{guimera07a}.

To quantify the plausability of our classification scheme, we analyze
the within-module consistency of metabolite pathway classification for
the top and the second levels of the metabolic network for {\it
E. coli} reconstructed at UCSD \cite{reed03}. For each module, we
first identify the pathways represented; then, we compute the fraction
of metabolites that are classified in the most abundant pathway.  We
find that there is a clear correlation between modules and known
pathways: At the top level, for all the modules except one, we find
that the most abundant pathway comprises more than 50\% of the
metabolites in the module. 

For the second level, we find that for most of the modules all the
metabolites are classified in the same pathway. We also detect smaller
pathways that are not visible at the top level (such as those for
polyketides and nonribosomal peptides, and for secondary metabolites).

Our results thus provide an objective description of cellular
metabolism that, while not affected by human subjectivity, captures our
current understanding of these networks. Interestingly, ``known''
pathways do not correspond to a single module at the top level,
implying that large pathways are in fact comprised of smaller
units. Intriguingly, these units are not necessarily uniform in
``pathway composition'' but are a mixture of sub-modules associated to
different pathways. Thus, an important question is how the modules we
identify relate to metabolism evolution\:\cite{raymond06}.

\begin{acknowledgments}
We thank U. Alon, A. Arenas, and S. Itzkovitz for providing network
data and W. Jiang for advice with the statistical
analysis. M.S.-P. and R.G. thank the Fulbright Program and the Spanish
Ministry of Education, Culture \& Sports. L.A.N.A. gratefully
acknowledges the support of the Keck Foundation, the J. S. McDonnell
Foundation and of a NIH/NIGMS K-25 award.
\end{acknowledgments}


\bibliographystyle{pnas}
\bibliography{/home/projects/BibTeX/ref-database,ref-specific}

\begin{thebibliography}{10}

\bibitem{anderson74}
Anderson, P.~W.
\newblock (1974) {\em Science} {\bf 177}, 393--396.

\bibitem{hartwell99}
Hartwell, L.~H, Hopfield, J.~J, Leibler, S,  \& Murray, A.~W.
\newblock (1999) {\em Nature} {\bf 402}, C47--C52.

\bibitem{pennisi05}
Pennisi, E.
\newblock (2005) {\em Science} {\bf 309}, 94.

\bibitem{oltvai02}
Oltvai, Z.~N \& Barab\'asi, A.-L.
\newblock (2002) {\em Science} {\bf 298}, 763--764.

\bibitem{alon03}
Alon, U.
\newblock (2003) {\em Science} {\bf 301}, 1866--1867.

\bibitem{itzkovitz05}
Itzkovitz, S, Levitt, R, Kashtan, N, Milo, R, Itzkovitz, M,  \& Alon, U.
\newblock (2005) {\em Phys Rev E Stat Nonlin Soft Matter Phys} {\bf 71},
  016127.

\bibitem{arenas06}
Arenas, A, D\'iaz-Guilera, A,  \& P\'erez-Vicente, C.~J.
\newblock (2006) {\em Phys. Rev. Lett.} {\bf 96}, 114102.

\bibitem{amaral04}
Amaral, L. A.~N \& Ottino, J.
\newblock (2004) {\em Eur. Phys. J. B} {\bf 38}, 147--162.

\bibitem{trusina04}
Trusina, A, Maslov, S, Minnhagen, P,  \& Sneppen, K.
\newblock (2004) {\em Phys. Rev. Lett.} {\bf 92}, 178702.

\bibitem{reichardt04}
Reichardt, J \& Bornholdt, S.
\newblock (2004) {\em Phys. Rev. Lett.} {\bf 93}, art. no. 218701.

\bibitem{ziv05}
Ziv, E, Middendorf, M,  \& Wiggins, C.~H.
\newblock (2005) {\em Phys Rev E Stat Nonlin Soft Matter Phys} {\bf 71},
  046117.

\bibitem{ravasz02}
Ravasz, E, Somera, A.~L, Mongru, D.~A, Oltvai, Z.~N,  \& Barab\'asi, A.-L.
\newblock (2002) {\em Science} {\bf 297}, 1551--1555.

\bibitem{soffer05}
Soffer, S.~N \& V\'azquez, A.
\newblock (2005) {\em Phys. Rev. E} {\bf 71}, art. num. 057101, 1--4.

\bibitem{guimera03}
Guimer\`a, R, Danon, L, D\'{\i}az-Guilera, A, Giralt, F,  \& Arenas, A.
\newblock (2003) {\em Phys. Rev. E} {\bf 68}, art. no. 065103.

\bibitem{clauset06}
Clauset, A, Moore, C,  \& Newman, M. E.~J.
\newblock (2006) {\em Structural inference of hierarchies in networks}.

\bibitem{pons06}
Pons, P.
\newblock (2006) Post-processing hierarchical community structures: Quality
  improvements and multi-scale view.
\newblock cs.DS/0608050.

\bibitem{newman04}
Newman, M. E.~J.
\newblock (2004) {\em Proc. Natl. Acad. Sci. USA} {\bf 101}, 5200--5205.

\bibitem{guimera05a}
Guimer\`a, R \& Amaral, L. A.~N.
\newblock (2005) {\em Nature} {\bf 433}, 895--900.

\bibitem{duch05}
Duch, J \& Arenas, A.
\newblock (2005) {\em Phys. Rev. E} {\bf 72}, art. no. 027104.

\bibitem{watts99}
Watts, D.~J.
\newblock (1999) {\em Small Worlds: The Dynamics of Networks between Order and
  Randomness}.
\newblock (Princeton University Press).

\bibitem{ye04}
Ye, Y \& Godzik, A.
\newblock (2004) {\em Genome Res} {\bf 14}, 343--53.

\bibitem{ihmels05}
Ihmels, J, Bergmann, S, Berman, J,  \& Barkai, N.
\newblock (2005) {\em PLoS Genet} {\bf 1}, e39.

\bibitem{newman04b}
Newman, M. E.~J \& Girvan, M.
\newblock (2004) {\em Phys. Rev. E} {\bf 69}, art. no. 026113.

\bibitem{clauset05}
Clauset, A.
\newblock (2005) {\em Phys. Rev. E} {\bf 72}, art. num. 026132, 1--6.

\bibitem{guimera04}
Guimer\`a, R, Sales-Pardo, M,  \& Amaral, L. A.~N.
\newblock (2004) {\em Phys. Rev. E} {\bf 70}, art. no. 025101.

\bibitem{guimera05}
Guimer\`a, R \& Amaral, L. A.~N.
\newblock (2005) {\em J. Stat. Mech.: Theor. Exp.} p. P02001.

\bibitem{fortunato07}
Fortunato, S \& Barthélemy, M.
\newblock (2007) {\em Proc Natl Acad Sci U S A} {\bf 104}, 36--41.

\bibitem{everitt01}
Everitt, B.~S, Landau, S,  \& Leese, M.
\newblock (2001) {\em Cluster Analysis}.
\newblock (Arnold Pub.).

\bibitem{schena02}
Schena, M.
\newblock (2002) {\em Microarray Analysis}.
\newblock (John Wiley and Sons, Inc.).

\bibitem{maslov02}
Maslov, S \& Sneppen, K.
\newblock (2002) {\em Science} {\bf 296}, 910--913.

\bibitem{guimera07}
Guimer\`a, R, Sales-Pardo, M,  \& Amaral, L. A.~N.
\newblock (2007) {\em Nature Phys.} {\bf 3}, 63--69.

\bibitem{kirkpatrick83}
Kirkpatrick, S, Gelatt, C.~D,  \& Vecchi, M.~P.
\newblock (1983) {\em Science} {\bf 220}, 671--680.

\bibitem{wasserman94}
Wasserman, S \& Faust, K.
\newblock (1994) {\em Social Network Analysis}.
\newblock (Cambridge University Press, Cambridge, UK).

\bibitem{koopmans75}
Koopmans, T \& Beckmann, M.
\newblock (1975) {\em Econometrica} {\bf 25}, 53--76.

\bibitem{tsafrir05}
Tsafrir, D, Tsafrir, I, Ein-Dor, L, Zuk, O, Notterman, D.~A,  \& Domany, E.
\newblock (2005) {\em Bioinformatics} {\bf 21}, 2301--2308.

\bibitem{schwarz78}
Schwarz, G.
\newblock (1978) {\em Ann. Stat.} {\bf 6}, 461--464.

\bibitem{akaike74}
Akaike, H.
\newblock (1974) {\em IEEE Transactions on Automatic Control} {\bf 19},
  716--723.

\bibitem{danon05}
Danon, L, D\'{\i}az-Guilera, A, Duch, J,  \& Arenas, A.
\newblock (2005) {\em J. Stat. Mech.: Theor. Exp.} p. P09008.

\bibitem{barrat04}
Barrat, A, Barth\'elemy, M, Pastor-Satorras, R,  \& Vespignani, A.
\newblock (2004) {\em Proc. Natl. Acad. Sci. USA} {\bf 101}, 3747--3752.

\bibitem{guimera04a}
Guimer\`a, R \& Amaral, L. A.~N.
\newblock (2004) {\em Eur. Phys. J. B} {\bf 38}, 381--385.

\bibitem{guimera05b}
Guimer\`a, R, Mossa, S, Turtschi, A,  \& Amaral, L. A.~N.
\newblock (2005) {\em Proc. Natl. Acad. Sci. USA} {\bf 102}, 7794--7799.

\bibitem{ma03}
Ma, H \& Zeng, A.-P.
\newblock (2003) {\em Bioinformatics} {\bf 19}, 270--277.

\bibitem{thiele05}
Thiele, I, Vo, T.~D, Price, N.~D,  \& Palsson, B.~{\O}.
\newblock (2005) {\em J. Bacteriol.} {\bf 187}, 5818--5830.

\bibitem{goto98}
Goto, S, Nishioka, T,  \& Kanehisa, M.
\newblock (1998) {\em Bioinformatics} {\bf 14}, 591--599.

\bibitem{kanehisa00}
Kanehisa, M \& Goto, S.
\newblock (2000) {\em Nucleic Acids Res.} {\bf 28}, 27--30.

\bibitem{reed03}
Reed, J.~L, Vo, T.~D, Schilling, C.~H,  \& Palsson, B.~{\O}.
\newblock (2003) {\em Genome Biol.} {\bf 4}, R54.

\bibitem{guimera07a}
Guimer\`a, R, Sales-Pardo, M,  \& Amaral, L. A.~N.
\newblock (2007) {\em Bioinformatics, in press}.

\bibitem{raymond06}
Raymond, J \& Segr\`e, D.
\newblock (2006) {\em Science} {\bf 311}, 1764--1767.

\bibitem{newman06}
Newman, M. E.~J.
\newblock (2006) {\em Proc. Natl. Acad. Sci. USA} {\bf 103}, 8577--8582.

\end{thebibliography}
\end{article}

\clearpage

\begin{table}
\begin{center}
\begin{tabular}{|l|c|c|c|c|}
\hline
Network & Size & Modules  & Main modules \\\hline
Air transportation& 3618 & 57 & 8 \\\hline 
Email & 1133 & 41 & 8\\\hline
Electronic circuit& 516 & 18 & 11\\\hline
{\em E. coli} KEGG & 739 & 39 & 13\\\hline
{\em E. coli} UCSD& 507 & 28 & 17\\\hline
\end{tabular}
\end{center}
%
\caption{Top-level structure of real-world networks. We show both the
total number of modules and the number of main modules at the top
level. Main modules are those comprised of more than 1\% of the
nodes. Note that there is no correlation between the size of the
network and the number of modules.
}
\label{t.realnetworks}
\end{table}
%
%
\end{document}